# Mechanical and Electronic Properties of Ferromagnetic Ga$_{1-x}$Mn$_x$As Using Ultrafast Coherent Acoustic Phonons

J. Qi,[1] J. A. Yan, [1] H. Park,[1] A. Steigerwald,[1] Y. Xu,[1] S. N. Gilbert, [1] X. Liu,[2] J. K. Furdyna,[2] S. T. Pantelides, [1] N. Tolk[1]

[1]Department of Physics and Astronomy, Vanderbilt University, Nashville, TN, 37235

[2]Department of Physics, University of Notre Dame, Notre Dame, IN 46556

Abstract

Ultrafast two-color pump-probe measurements, involving coherent acoustic phonon (CAP) waves, have provided information simultaneously on the mechanical properties and on the electronic structure of ferromagnetic Ga$_{1-x}$Mn$_x$As. The elastic constant C$_{11}$ of Ga$_{1-x}$Mn$_x$As (0.03 ≤x≤ 0.07) are observed to be systematically smaller than those of GaAs. Both C$_{11}$ and V$_s$ of Ga$_{1-x}$Mn$_x$As are found to increase with temperature (78 K ≤T≤ 295 K), again in contrast to the opposite behavior in GaAs. In addition, the fundamental bandgap (at $E_0$ critical point) of Ga$_{1-x}$Mn$_x$As is found to shift slightly to higher energies with Mn concentration.

# I. Introduction

The introduction of carrier-mediated III-Mn-V ferromagnetic semiconductors opens up promising opportunities to combine semiconducting properties and robust magnetism into conventional electrical and optical devices [1, 2]. In the process of explaining recent experimental observations, along with attempts to develop a theory of hole-mediated ferromagnetism in "prototypical" ferromagnetic GaMnAs systems, the relationship between mechanical, magnetic and electronic properties has been extensively explored [2-4]. For instance, much attention has been given to understanding of the effect of strain on the magnetic easy axis and on anisotropy fields arising from the substrate-film lattice mismatch [5-8]. Specifically, however, when constructing a theoretical model, due to the lack of corresponding experimental values for the GaMnAs system, values of elastic constants in the strain-associated Hamiltonian of GaMnAs are generally assumed to be the same as those of GaAs [7-9]. On the other hand, over the past several years the nature of the electronic structure in carrier-mediated ferromagnetism of $Ga_{1-x}Mn_xAs$ has been the subject of an important and lively debate [2, 3]. One effective way to reveal the band structure of this material is to study the evolution of the critical points in energy by measuring relevant optical constants such as the complex refractive index [ $\hat{n} = n + ik$ ] or the complex dielectric constant [ $\hat{\varepsilon} = \varepsilon_1 + i\varepsilon_2$ ] [3]. Recently Burch et al. [10] used spectroscopic ellipsometry and observed a large blue-shifting of the critical point $E_1$ with Mn doping. They attributed this to a Mn-induced impurity band (IB) existing in this material system and this IB has strongly hybridized with the GaAs valence band (VB) near the $L$-point. However, they did not observe any changes for all other critical points as Mn concentration varies.

Although many time-domain studies of carrier and spin dynamics have been done on (III,Mn)V structures over the years, few experiments are reported to measure the mechanical properties and reveal the electronic structure of this system using pump-probe spectroscopy. In this article, we report measurements of the elastic constant $C_{11}$ utilizing an ultrafast pump-probe approach involving coherent acoustic phonons (CAPs) propagating in ferromagnetic $Ga_{1-x}Mn_xAs$ ($0.03 \leq x \leq 0.07$) systems. The measured values are found to be very different from those of GaAs and low-temperature grown GaAs (LT-GaAs). These observations are partially explained by density functional calculations. Experimentally, we also observe a clear blueshift of the critical point $E_0$ (fundamental bandgap) with Mn doping in $Ga_{1-x}Mn_xAs$ for the first time.

## II. Samples and Experimental Setup

Our ferromagnetic $Ga_{1-x}Mn_xAs$ sample is grown by the low temperature molecular beam epitaxy (LT-MBE) method. Before the growth of the $Ga_{1-x}Mn_xAs$ (x=0.03, 0.05, and 0.07) layer with a thickness 1 μm, a low temperature (~270 °C) GaAs (LT-GaAs) buffer layer with a thickness 300 nm was grown on a semi-insulating GaAs(001) substrate. For the purpose of comparison, one GaAs(001) substrate and one LT-GaAs sample (with no Mn) with a layer thickness of 1.3 μm grown under similar conditions were also used in the experiment. The ferromagnetic $Ga_{1-x}Mn_xAs$ samples with x=0.03, 0.05 and 0.07 have Curie temperatures around 35 K, 55 K, and 53 K, respectively.

In our setup, we used a two-color all optical pump-probe technique. The pump photon energies were around 3.1 eV (~400 nm), significantly above the fundamental bandgap $E_0$ of GaAs (1.42 eV (~873 nm) at 295 K and 1.51 eV (~821 nm) at 78 K). The probe light, on the other hand, is chosen to be near $E_0$. Such high pump photon energy leads to the creation of coherent phonons generated near the surface of the $Ga_{1-x}Mn_xAs$ layer arising from the femtosecond excitation light pulse. The subsequent CAP wave (strain pulse), which alters locally the optical properties (the complex refractive index), propagates into the $Ga_{1-x}Mn_xAs$/GaAs(001) and is detected by monitoring the reflected probe pulse. Such CAPs have been widely observed using pump-probe spectroscopic technique in many material systems [11-16]. All the following experiments measuring transient reflectivity changes $\Delta R/R$ were performed employing a Ti:Sapphire laser with a repetition rate of 76 MHz, which produces ~150 fs wide pulses in the wavelength range between ~720nm to ~950nm. The temperature can be varied between 78 K and 295 K. Both pump and probe beams (typical intensity ratio of 8:1) were focused onto the sample at the same spot with a diameter of around 100 μm. The pump light typically had a fluence of 1 μJ/cm$^2$.

## III. Results

Figure 1 shows the transient reflectivity signal $\Delta R/R$ measured in the $Ga_{0.97}Mn_{0.05}As/GaAs$ sample at 78 K. The inset is the measured oscillatory response obtained by subtracting the relaxation background. The probe pulse is centered at 830 nm, with photon energy somewhat below the fundamental band gap of GaAs (~821 nm). It can be seen that the total response consists of a fast transient (on the order of a few picoseconds) followed by a tail superimposed with two distinct damped oscillatory regimes, separated by a narrow transition region. The initial fast transient is typical of electronic contribution to the pump-probe signal.

The observed oscillations may be explained by a propagating strain pulse model [11], where the oscillations originate from the interference of probe beams reflected from the top sample surface and the propagating strain pulse. The reflectivity change $\Delta R/R$ for the oscillatory behavior may be written [11]

$$\Delta R/R \propto Ae^{-t/\tau}\cos(2\pi t/T_p + \varphi) \qquad (1)$$

where A is the amplitude, $\tau$ is the damping time, $\varphi$ is phase shift, and $T_p$ is the oscillation period given by $T_p = \lambda/(2nV_s\cos\theta)$. Here n is the refractive index, $V_s$ is speed of sound, and $\theta$ (~0 in our experiments) is the angle of incidence of the probe light in GaMnAs with respect to the normal to sample surface. In the 'near-field' approximation [5], the damping time $\tau$ is further related to the absorption properties of the material by $\tau = 1/(2\alpha V_s) = \lambda(4\pi V_s\kappa)^{-1}$, where $\alpha$ is the absorption coefficient and $\kappa$ is the imaginary part of the complex refractive index $\tilde{N}(=n+i\kappa)$. The amplitude A is, furthermore, connected to the change of the local complex refractive index through the z-component strain ($\eta_{33}$) by the relation $A \propto |\delta\tilde{N}/\delta\eta_{33}|$ [11, 14].

The generated CAP wave first travels through $Ga_{1-x}Mn_xAs$ normal to the surface at the speed

$V_{s(GaMnAs)}$ of the longitudinal acoustic phonon (LAP) before it reaches the LT-GaAs and GaAs regions. After passing through the $Ga_{1-x}Mn_xAs$/LT-GaAs interface, it continuously propagates into the LT-GaAs and GaAs layers with LAP speed $V_{s(LT-GaAs)}$ and $V_{s(GaAs)}$, respectively. We estimate that it takes about $\Delta t \approx 200$ ps for the strain pulse to arrive at LT-GaAs layer assuming that $V_{s(GaMnAs)}$ is approximately equal to $V_{s(GaAs)}$ (~4.78×$10^3$ m/s at 78 K [17]). This can be seen from the experimental results shown in Fig. 1.

We investigated the dependence of the oscillations near the bandgap of GaAs on the probe wavelength. The oscillations induced by CAP waves traveling in the $Ga_{1-x}Mn_xAs$ layer decay markedly at all the wavelengths studied. In contrast, the oscillations in the GaAs response persist for a very long time at probe wavelengths below the bandgap of GaAs. Applying Eq. (1) to the studied material, we can numerically fit our experimental data at different wavelengths, and obtain significant physical parameters, including the period, damping time, and amplitude of the oscillations. We note that the two-color pump-probe technique enables us to get the oscillation profiles with a high signal/noise ratio. Standard errors of parameters $T_p$ and $\tau$ of the oscillation are less than 1% and 3%, respectively.

As is well known, the speed of LAP waves propagating along [001] is given by $V_s=(C_{11}/\rho)^{1/2}$ for cubic-symmetry solids with elastic constant $C_{11}$ and density $\rho$. Therefore, using this equation and the expression $T_p=\lambda/(2nV_s)$, we can determine two of the important mechanical properties, the LAP speed $V_s$ along [001] and the corresponding elastic constant $C_{11}$, which have never before been reported for the ferromagnetic $Ga_{1-x}Mn_xAs$ system. Here, we also report temperature (*T*) dependent measurements of $C_{11}$.

The refractive indices *n* of GaAs, LT-GaAs and $Ga_{1-x}Mn_xAs$ at room temperature (T=295 K) are directly measured by spectroscopic ellipsometry. In principle, refractive index $n$ depends on both temperature T and wavelength $\lambda$. In this study, at low temperatures we used $n$ measured at T=295 K with shorter wavelengths (i.e., $\lambda \leq 800$ nm) away from the fundamental bandgap $E_0$ to obtain $V_s$. It means that we assume $n$(T, $\lambda \leq 800$ nm) is independent of T for 78 K$\leq$T$\leq$295 K. This assumption (or approximation) is based on two considerations [18, 19]: (1) in the infrared non-resonant absorption regime, refractive indices $n$ for both GaAs and GaMnAs change slightly as temperature varies (no more than 1% for 78 K$\leq$T$\leq$295 K); (2) $n^{-1}dn/dT$ has relatively large values at the critical energy points (i.e., $E_0$). However, validity of such assumption needs to be carefully checked.

Firstly, the calculated propagating time of the strain pulse inside different layer using the obtained $V_s$ based on above assumption and the known film thickness should be consistent with the value directly obtained by the oscillation amplitude in the reflectivity measurements. Our previous works [15, 20] suggest that tracing the oscillation amplitude changes across different layers provides a very powerful tool in depth profiling and material characterizing of heterostructure semiconductors. The traveling time in different layers can be accurately acquired by derivatives of the envelope of the oscillations. It should be pointed out that the instantaneous created strain pulse here has a spatial width, which is around 100 nm and is centered on ~50 nm away from the top surface determined by the absorption depth of pump light. Thus, for Ga$_{0.95}$Mn$_{0.05}$As layer at T=78 K, according to the value of $V_s$=4655 m/s and the effective propagating distance ~950 nm, the time can be estimated to be $t_1$~204 ps. It agrees very well with the value $t_2$~205 ps derived from the oscillation profile, which is indicated in Fig. 1(b). Following the same procedure, we can get the relative change $|t_1-t_2|/t_1$ is less than 0.5% at different temperatures on different samples. Therefore, our assumption above is well supported by the experiments.

We note that the obtained speed of sound $V_s$ has to be independent of wavelength $\lambda$ for $\lambda \leq 800$ nm.

Our results between 78 K and 295 K for 760 nm$\leq\lambda\leq$800 nm show that $|\Delta V_s|/V_s$ is less than 0.5%, which is quite small and confirms $V_s$ constant. Therefore, reasonable $V_s$ can be obtained at any wavelength between 760 nm and 800 nm. Here, average value of $V_s$ at two wavelengths (760 nm and 800 nm) is chosen, and employed to calculate $C_{11}$.

In zinc-blende GaAs, the thermal expansion effect has been measured to be very small in the temperature range of 50~300 K. The typical thermal expansion coefficient value is $\alpha_{th} = \frac{1}{a_0}\frac{da_0}{dT} = 6\times10^{-6}\,K^{-1}$ [18]. For $Ga_{1-x}Mn_xAs$ (x=0.01-0.09), it is believed that the temperature dependence of lattice constants are almost the same as that in bulk GaAs, and the relaxed lattice constant for $Ga_{1-x}Mn_xAs$ follows a=0.566+(0.598-0.566)x nm [1,2]. Assuming the primary lattice location of Mn to be substitutional $Mn_{Ga}$, and up to ~20% of Mn as residing in interstitial positions [2], the density of as-grown $Ga_{1-x}Mn_xAs$ sample is expected to be roughly the same as that of GaAs. In real calculation of $C_{11}$, at 295 K, $\rho$ = 5.317 g/cm$^3$ was assumed for GaAs and LT-GaAs, and $\rho$ = 5.317×(1-0.169x) g/cm$^3$ was for $Ga_{1-x}Mn_xAs$. The thermal expansion was corrected in the same way using $\alpha_{th}$=6×10$^{-6}$ $K^{-1}$. In fact, relative change of $C_{11}$ due to thermal expansion will not exceed 0.4% by a simple estimation using the above $\alpha_{th}$ value.

Figure 2 displays the experimental results of $T_p(\lambda$=800 nm), and the corresponding $C_{11}$. Our measured results for GaAs agree quite well with previously reported values [17, 18]. It can also be seen that GaAs and LT-GaAs have very similar $C_{11}$ in the studied temperature regime. This indicates that the disordering effect caused by the large number of $As_{Ga}$ antisite defects (with a concentration ~10$^{19}$ cm$^{-3}$) in LT-GaAs plays a minor role in the mechanical properties along [001]. Figure 2 demonstrates that $C_{11}$ measured for the $Ga_{1-x}Mn_xAs$ (x = 0.03, 0.05 and 0.07) samples are also close to each other, and hence have similar mechanical properties. Surprisingly, we found that $C_{11}$ increase with $T$ ($\delta C_{11}/\delta T > 0$) in $Ga_{1-x}Mn_xAs$ samples, in striking contrast to their behavior in both GaAs and LT-GaAs, where they are observed to decrease as the temperature increases ($\delta C_{11}/\delta T < 0$). A linear fitting of $C_{11}$ was carried out for GaAs and $Ga_{0.97}Mn_{0.03}As$, respectively. The fitting yields $C_{11}(GaAs)=121.2-6.09\times10^{-3}T$ ($10^{10}$ dyn/cm$^2$) and $C_{11}(Ga_{0.97}Mn_{0.03}As)=113.3+8.68\times10^{-3}T$ ($10^{10}$ dyn/cm$^2$). It can also be seen that the obtained $C_{11}$ for the ferromagnetic $Ga_{1-x}Mn_xAs$ system are systematically smaller than for GaAs or LT-GaAs in the studied temperature range. Specifically, the relative change $|C_{11(GaAs)} - C_{11(GaMnAs)}|/ C_{11(GaMnAs)}$ increases from ~3% at 295 K to ~6% at 78 K. In fact, $V_s$ shows similar behavior as that of $C_{11}$.

## IV. Discussions

To understand the decrease of $C_{11}$ in $Ga_{1-x}Mn_xAs$, we performed spin-polarized density functional calculations of the elastic constants for x = 3.1 % within the generalized gradient approximation (GGA) using the VASP code [21]. Ultrasoft pseudopotentials for Ga, Mn, and As were employed. The plane-wave cutoff energy was set to 450 eV. For the elastic constants $C_{11}$ and $C_{12}$, we follow the method proposed by Nielsen and Martin [22], while $C_{44}$ is obtained according to Mehl et al. [23]. The calculated results are listed in Table 1. Despite the absolute deviation from the experimental values, the $C_{11}$ (93.4 GPa) of $Ga_{1-x}Mn_xAs$ is found to decrease by 8.6 % at 0 K compared to that (102.2 GPa) of GaAs, consistent with the experimental observation (decrease by ~3% at 295 K and ~6% at 78 K). In Figs. 3(a) and 3(b), we further show the charge density difference for the strained and unstrained bulk GaAs and $Ga_{1-x}Mn_xAs$, respectively. The charge densities of uniaxially strained crystals reveal interesting information on how electrons as well as nuclear positions respond to macroscopic strain. A small strain $\varepsilon_1 = -0.01$ is applied along [100] in GaAs and $Ga_{1-x}Mn_xAs$, respectively. As shown in Fig. 3(b), the substitutional $Mn_{Ga}$ doping induces an evident charge accumulation near the Mn nuclei, indicating that Mn-As bond becomes more ionic. Moreover, the effect of Mn doping is rather long-range: charge distributions are significantly distorted up to the third neighbors compared to GaAs.

The above calculations combined with the phenomenological Keating model [24] may give a more understandable picture. According to Keating [24], $C_{11} = \alpha (1+3\beta/\alpha)/4a_0$, where $\alpha$ is the central (or bond-stretching) force constant, $\beta$ is the mean value of the noncentral (or bending) force constant, and $a_0$ is the lattice constant. Since the lattice constant varies only slightly with substitutional $Mn_{Ga}$ doping, the contribution from $a_0$ of the as-grown $Ga_{1-x}Mn_xAs$ ($0.03\leq x\leq0.07$) is expected to be negligible (<0.4%). On the other hand, the ionicity of a material can be characterized by the ratio $\beta/\alpha$ [25]. Ideally, the stronger the ionicity, the smaller the quantity $\beta/\alpha$ [25]. Therefore, $\beta/\alpha$ in the presence of Mn is expected to decrease compared to GaAs. This illustrates why the elastic constants in $Ga_{1-x}Mn_xAs$ become smaller than those in GaAs.

The temperature dependence of elastic constant $C_{11}$ in GaAs conforms to theoretical expectation. Namely, from 78 K to 295 K, the elastic constant $C_{11}$ almost decreases linearly with increasing temperature [26, 27]. In contrast, $C_{11}$ of $Ga_{1-x}Mn_xAs$ shows peculiar temperature behavior and increases with increasing T, as shown in Fig. 2(b). Apparently, upon doping of small fraction of Mn, the elastic properties of GaAs have been altered dramatically. Although the underlying mechanism is not yet fully understood at this moment, we speculate that it might be analogous to similar phenomena observed in some transition metals: the shear elastic constants $C_{44}$ of Nb, V, Ti and some alloy systems were observed to show anomalous temperature behavior nearly thirty years ago [28]. The origin has been ascribed to unique features of the Fermi surface in these systems. In $Ga_{1-x}Mn_xAs$, the possible magnetic-elastic coupling induced by Mn may even complicate the situation and calls for further theoretical and experimental efforts.

In our experiments, since $T_p \propto 1/n$ and $\tau \propto 1/k$, systematic analysis of $T_p$ and $\tau$ as function of wavelength promises to provide some important insights into the electronic structure of the $Ga_{1-x}Mn_xAs$ system. Figs. 4 and 5 show our experimental oscillation period $T_p$ and decay time $\tau$ in terms of different probe wavelength $\lambda$ for $Ga_{1-x}Mn_xAs$, LT-GaAs, and GaAs measured at 78 K. As can be seen from Fig. 4, the oscillation periods are close to linear as a function of probe light wavelength for GaAs. This behavior agrees with the expression $T_p=\lambda/(2nV_s)$. In addition, the profile of $T_p(\lambda)$ for LT-GaAs is very similar to that for GaAs. This indicates that both mechanical and optical properties of these two materials are close, or that the As antisite defects play no strong role in determining $V_s$ and $n$. However, the $T_p(\lambda)$ curves for $Ga_{1-x}Mn_xAs$ ($0.03 \leq x \leq 0.07$) differ greatly from those for GaAs or LT-GaAs. This implies that there is a big difference in $V_s$ and $n$ between $Ga_{1-x}Mn_xAs$ and GaAs or LT-GaAs. Such differences have already been confirmed by our discussion above as well as by previous optical measurements [3, 10]. Also, in $Ga_{1-x}Mn_xAs$ samples we find that the $T_p(\lambda)$ curve begins to strongly deviate from linear dependence in the long wavelength regime (>850 nm). This observation has to be associated with a distinct nonlinear $n(\lambda)$ dependence for wavelengths corresponding to energies below the GaAs fundamental bandgap $E_0$. The disordering effects are known to change optical properties (e.g. $n$ and $\kappa$) in the material [3, 10]. Therefore, comparison of $T_p$ in $Ga_{1-x}Mn_xAs$ and LT-GaAs indicates that such strong nonlinearity is mainly due to Mn impurities rather than the As antisite defects.

In Fig. 5, a sharp change in GaAs profile can be seen around 1.51 eV (~821 nm indicated by the dash line), which is the fundamental bandgap (the critical point $E_0$ in energy). This kind of feature becomes diminished in LT-GaAs, and disappears in the $Ga_{1-x}Mn_xAs$ systems. It is consistent with previous studies on $Ga_{1-x}Mn_xAs$ [3, 10], where the bandgap is broadened due to the $As_{Ga}$ antisites and Mn impurities.

A closer look at the details of the damping time $\tau(\lambda)$ in Fig. 5, we observe the following important features. The structure of $\tau(\lambda)$ gradually evolves from a sharp step-like feature near $E_0$ in the GaAs data into a multi-peak structure in $Ga_{1-x}Mn_xAs$. Surprisingly, the multi-peak structure clearly blue-shifts with Mn doping. This finding implies a blue-shifting of the fundamental bandgap as Mn concentration increases. Such shifting has never been directly observed before, although magnetic circular dichroism (MCD) experiments suggest that $E_0$ blueshifts slightly with doping [29]. Firstly, the compressive strain in the $Ga_{1-x}Mn_xAs$ film cannot account for the blueshift of the critical point $E_0$ [10]. In order to unveil this observation, one needs to understand the electronic structure of the ferromagnetic $Ga_{1-x}Mn_xAs$. It should be pointed out that there are two scenarios proposed to describe this system: the debate is whether the Fermi energy sits inside the GaAs heavy and light hole VB or inside an Mn induced IB above the top of GaAs VB [3]. However, we cannot explain the observed blueshift using the VB scenario, since Zhang and Sarma theoretically predicted on the basis of the VB scenario that as x increases the bandgap becomes smaller resulting in a red-shifting in the $Ga_{1-x}Mn_xAs$ system [30]. Thus, our results indicate that the VB scenario does not adequately describe the Mn-doped GaAs band structure. This blue-shifting should be associated with the Mn-induced IB. As is known, the *p-d* hybridization exists between Mn-induced IB and GaAs VB. Such hybridization is *k*-dependent and very strong near the *L*-point, which explains the observed large blue-shift of the critical point $E_1$ in Ref. [10]. Therefore, although the *k*-dependent *p-d* hybridization is expected to be small around the $\Gamma$ point, our experiments indicate that the small $V_{pd}$ still can lead to the observable blue-shifting of the fundamental bandgap as long as the experimental technique is sensitive enough to reveal the subtle feature.

## V. Conclusions

In conclusion, employing femtosecond two-color pump-probe spectroscopy we have completed investigations on pronounced oscillatory behavior in the reflectivity curves of magnetic semiconductor (Ga,Mn)As systems. We report experimental value of the elastic constant $C_{11}$ on ferromagnetic $Ga_{1-x}Mn_xAs$ samples. Both $V_s$ and $C_{11}$ in the (Ga,Mn)As systems are systematically smaller than in GaAs at temperature between 78 K and 295 K. The temperature dependence of $C_{11}$ for $Ga_{1-x}Mn_xAs$ and GaAs shows totally different behavior. The slight blueshifting of the $E_0$ critical point with Mn doping in $Ga_{1-x}Mn_xAs$ system was directly observed, and is attributed to the hybridization of the Mn IB with the GaAs VB. This work was supported at Vanderbilt by DOE through Grant DE-FGO2-99ER45781 and at Notre Dame by NSF Grant DMR06-03762. J. Yan and S. T. Pantelides acknowledge the McMinn Endowment at Vanderbilt University.

Figure Captions

Figure 1 (a) Pump-probe response of $Ga_{0.95}Mn_{0.05}As/LT\text{-}GaAs/GaAs$ at 78 K for probe wavelength 830 nm; (b) Derivatives of envelope of oscillations $|(\Delta R/R)_{env}|$. Inset: the subtracted oscillatory response. The regimes determined by the GaMnAs and LT-GaAs layers are indicated by horizontal arrows in the inset.

Figure 2 (a) Oscillation period $T_p$ at 800 nm and (b) elastic constant $C_{11}$ as a function of temperature T for $Ga_{1-x}Mn_xAs$, LT-GaAs and GaAs. Solid lines are a linear fit for GaAs and $Ga_{0.97}Mn_{0.03}As$, respectively.

Figure 3 (Color online) Charge density difference $\Delta\rho$ for (a) bulk GaAs and (b) $Ga_{1-x}Mn_xAs$ (x = 3.1%) in the {110} plane which contains a zigzag atomic chain. $\Delta\rho$ is defined as the charge density of a strained crystal minus the charge density of an unstrained crystal. The applied strain is $\varepsilon_1 = -0.01$. Contour steps are $4\times10^{-4}$ e/$\text{Å}^3$, and red solid (blue dashed) contours indicate charge accumulation (depletion).

Figure 4 Oscillation period $T_p$ as a function of the wavelength $\lambda$ of probe light for $Ga_{1-x}Mn_xAs$, LT-GaAs and GaAs at 78 K. Dashed line represents the position of the critical point $E_0$ of GaAs.

Figure 5 (a) Damping time $\tau$ as a function of the wavelength $\lambda$ of probe light for $Ga_{1-x}Mn_xAs$, LT-GaAs and GaAs at 78 K. Dashed line represents the position of the critical point $E_0$ of GaAs. In (b), curves for $Ga_{1-x}Mn_xAs$ are shifted downward with increasing x for clarity of the blueshift.

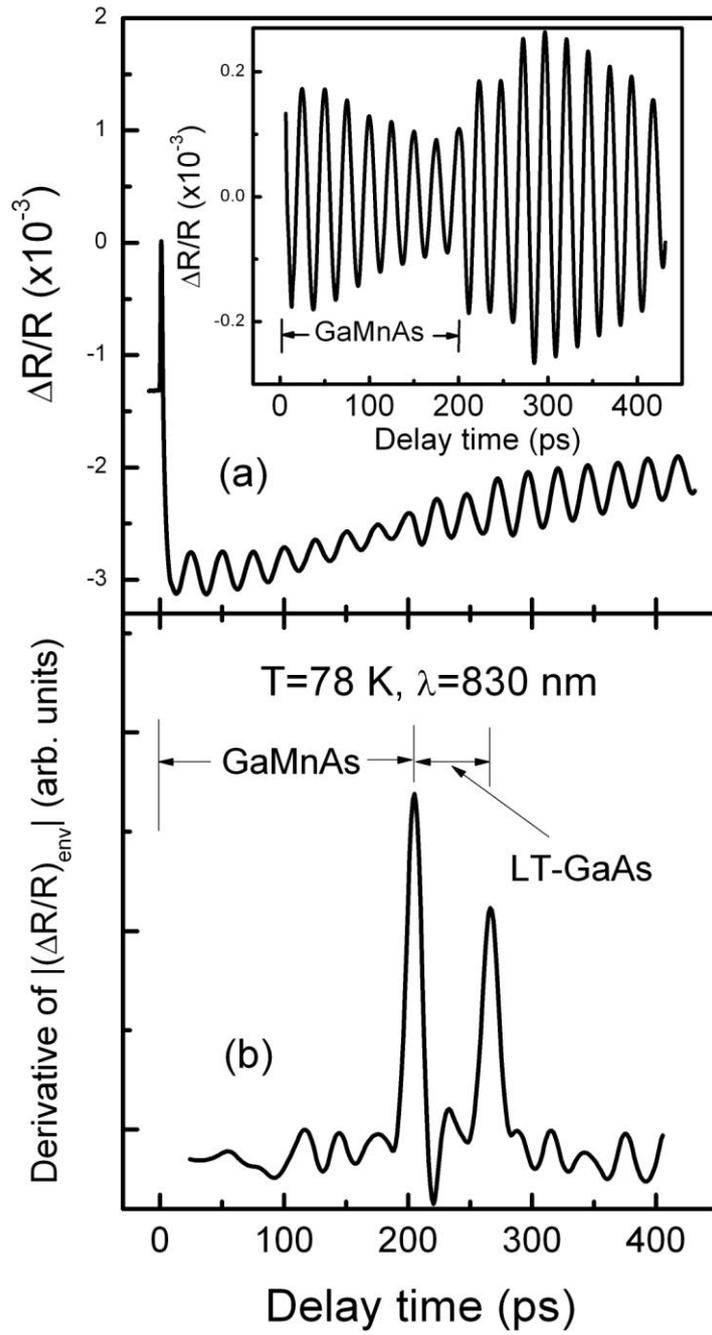

Figure 1

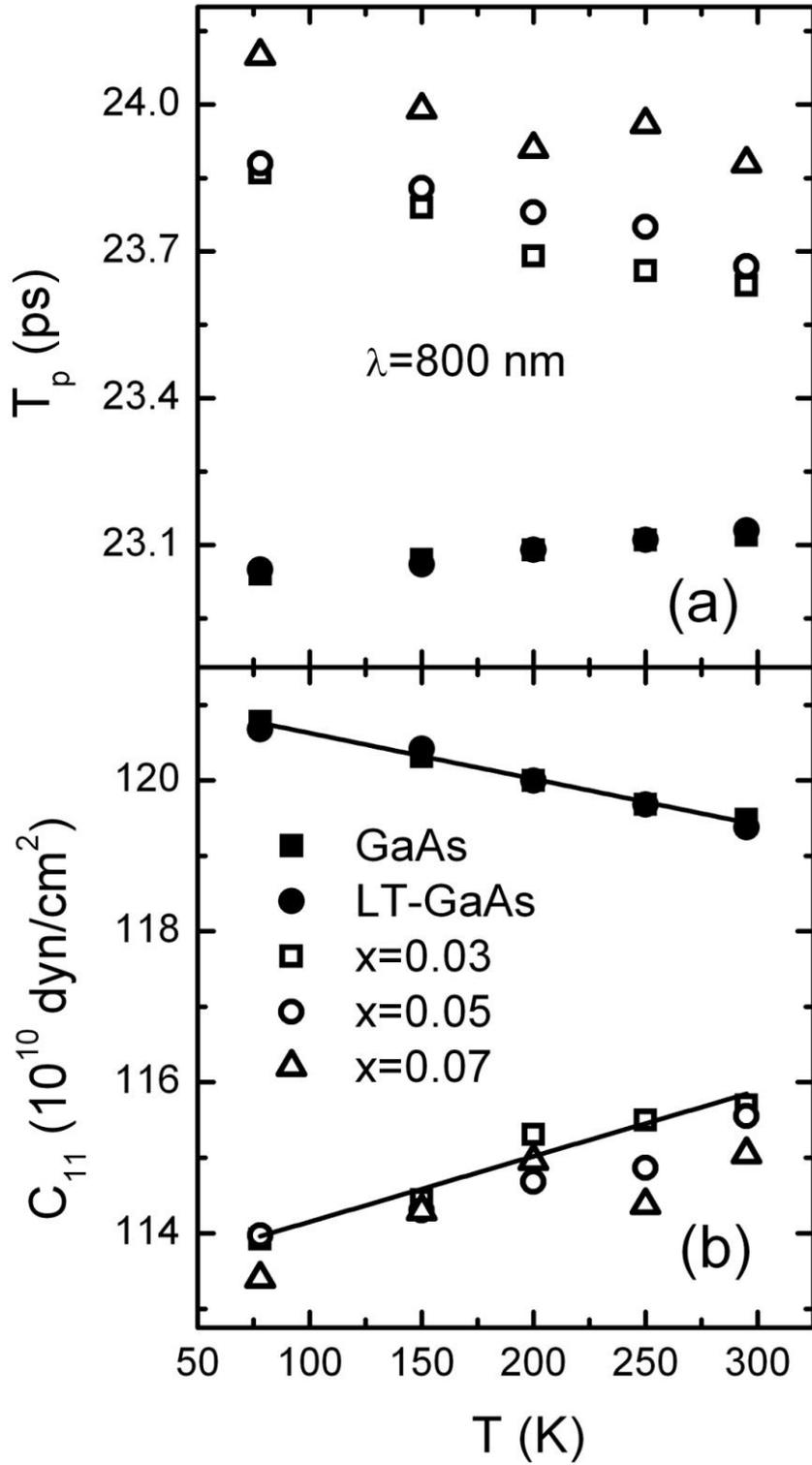

Figure 2

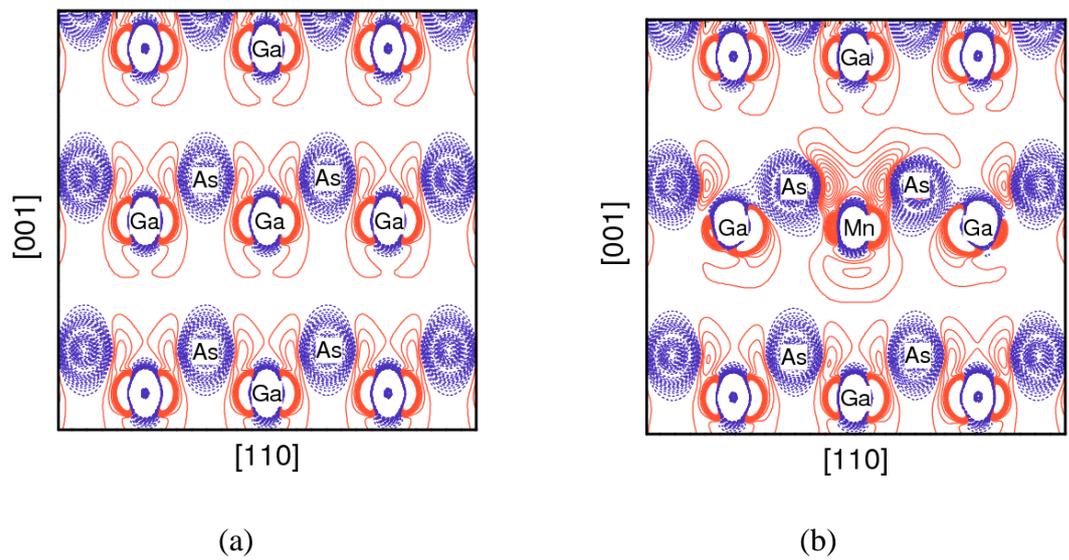

(a)                                           (b)

Figure 3

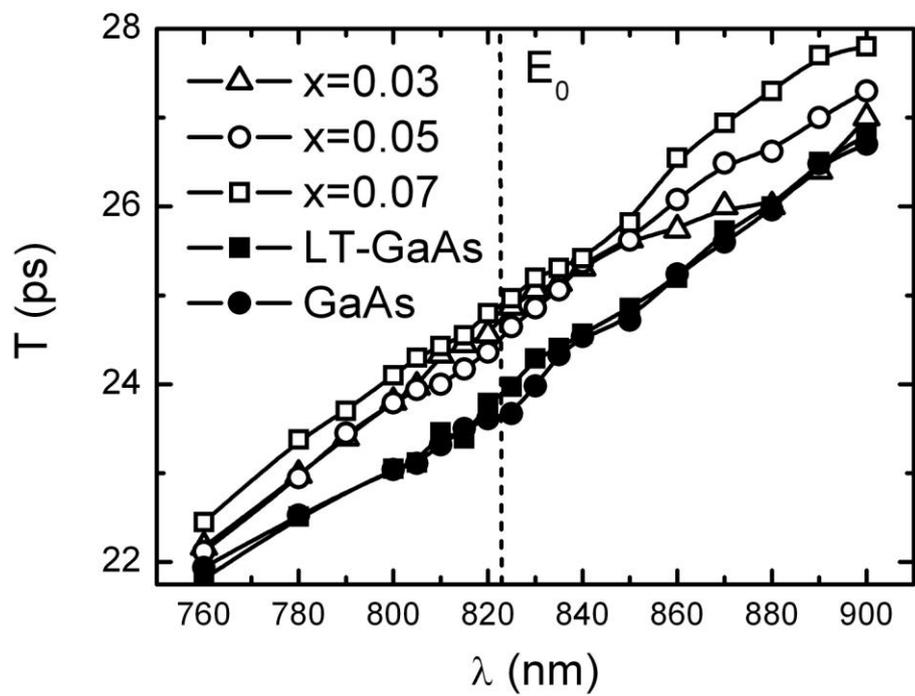

Figure 4

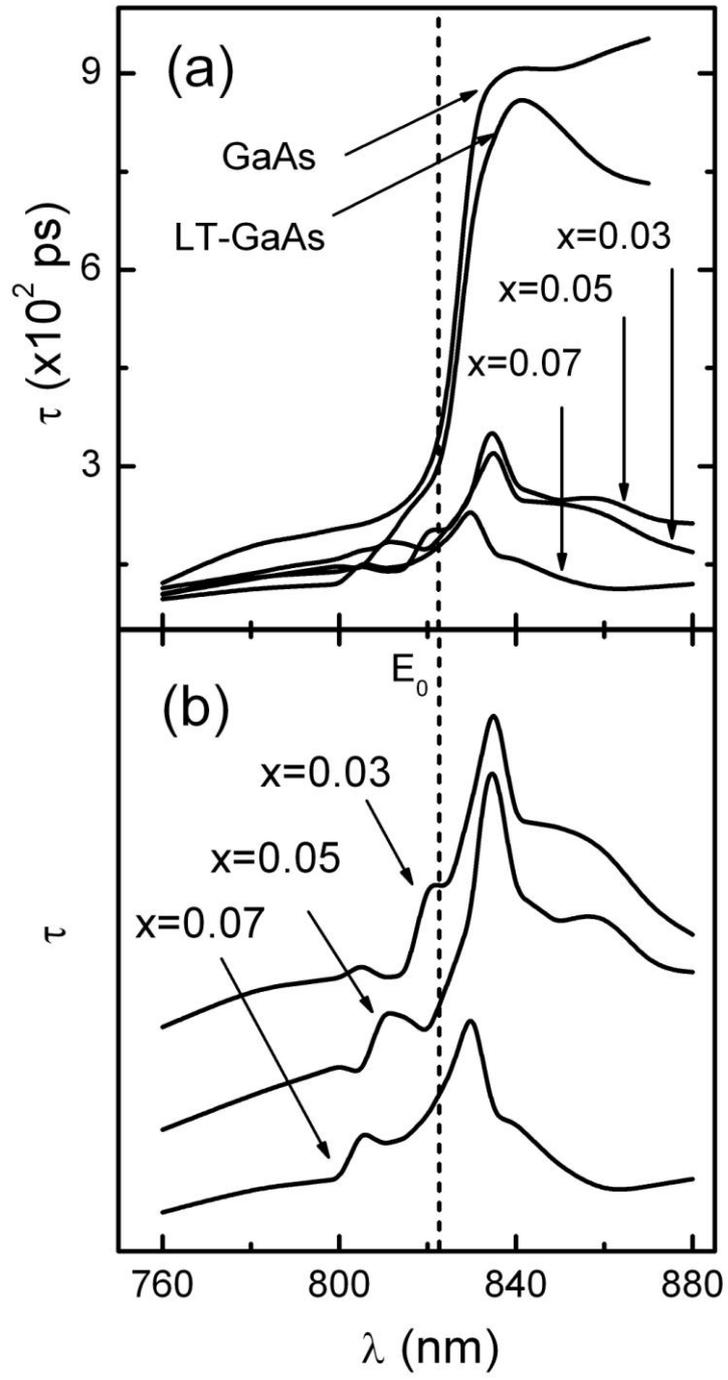

Figure 5

Table 1 Calculated elastic constants of bulk GaAs, zince-blende bulk MnAs, and $Ga_{1-x}M_xAs$ (x = 0.03).

| | GaAs | MnAs (zinc-blende) | $Ga_{1-x}Mn_xAs$ (x=0.03) |
|---|---|---|---|
| $C_{11}$ (GPa) | 102.2 | 62.0 | 93.4 |
| $C_{12}$ (GPa) | 41.4 | 59.1 | 45.1 |
| $C_{44}$ (GPa) | 56.0 | 2.2 | 47.6 |
| $B=(C_{11}+2C_{12})/3$ (GPa) | 61.7 | 60.2 | 61.22 |